\documentstyle[12pt,aaspp4]{article}
\input epsf
\lefthead{H. Netzer, T.J. Turner}
\righthead{NGC 1068}
\begin{document}
\title{Soft X-ray lines and gas composition in  NGC 1068}
\centerline{ \today}
\author {Hagai Netzer \altaffilmark{1},
       T.J.Turner,\altaffilmark{2},\altaffilmark{3} }
\altaffiltext{1}{School of Physics and Astronomy and the Wise
Observatory, The Raymond and  Beverly Sackler Faculty of Exact Sciences,
Tel-Aviv University, Tel-Aviv 69978, Israel.}
\altaffiltext{2}{Laboratory for High Energy Astrophysics,
  Code 660, NASA/Goddard Space Flight Center, Greenbelt, MD 20771}
\altaffiltext{3}{University Space Research Association}
%
%
% following is a copy of macros.tex
%
\def\etal{{\it et al.}}
\def  \vs          {{\it vs.} }
\def  \eg          {{\rm e.g.}}
\def\G{$\Gamma$}
\def\grs{$\gamma$-rays}
\def\egs{{\it EGRET}~}
\def\gr{$\gamma$-ray}
\def\IUE{{\it IUE}}

%
% lines
%
\def  \La          {\ifmmode {\rm Ly}\alpha \else Ly$\alpha$\fi}
\def  \Ka          {\ifmmode {\rm K}\alpha \else K$\alpha$\fi}
\def  \Lb          {\ifmmode {\rm L}\beta \else L$\beta$\fi}
\def  \Ha          {\ifmmode {\rm H}\alpha \else H$\alpha$\fi}
\def  \Hb          {\ifmmode {\rm H}\beta \else H$\beta$\fi}
\def  \Pa          {\ifmmode {\rm P}\alpha \else P$\alpha$\fi}
\def  \HeI        {\ifmmode {\rm He}\,{\sc i}\,\lambda5876
                     \else He\,{\sc i}\,$\lambda5876$\fi}
\def  \HeII        {\ifmmode {\rm He}\,{\sc ii}\,\lambda4686
                     \else He\,{\sc ii}\,$\lambda4686$\fi}
\def  \HeIIa        {\ifmmode {\rm He}\,{\sc ii}\,\lambda1640
                     \else He\,{\sc ii}\,$\lambda1640$\fi}
\def  \HeIIb        {\ifmmode {\rm He}\,{\sc ii}\,\lambda1085
                     \else He\,{\sc ii}\,$\lambda1085$\fi}
\def  \CII        {\ifmmode {\rm C}\,{\sc ii}\,\lambda1335
                     \else C\,{\sc ii}\,$\lambda1335$\fi}
\def  \CIII        {\ifmmode {\rm C}\,{\sc iii}\,\lambda977
                     \else C\,{\sc iii}\,$\lambda977$\fi}
\def  \CIIIb       {\ifmmode {\rm C}\,{\sc iii]}\,\lambda1909
                     \else C\,{\sc iii]}\,$\lambda1909$\fi}
\def  \CIV         {\ifmmode {\rm C}\,{\sc iv}\,\lambda1549
                     \else C\,{\sc iv}\,$\lambda1549$\fi}
\def  \bOIIIb       {\ifmmode {\rm [O}\,{\sc iii]}\,\lambda5007
                     \else [O\,{\sc iii]}\,$\lambda5007$\fi}
\def  \OIIIb       {\ifmmode {\rm O}\,{\sc iii]}\,\lambda1663
                     \else O\,{\sc iii]}\,$\lambda1663$\fi}
\def  \OVIII        {\ifmmode {\rm O}\,{\sc viii}\,653~{\rm eV}
                     \else O\,{\sc viii}\,$653~{\rm eV}$\fi}
\def  \OVII        {\ifmmode {\rm O}\,{\sc vii}\,568~eV
                     \else O\,{\sc vii}\,$568~{\rm eV}$\fi}
\def  \OVI         {\ifmmode {\rm O}\,{\sc vi}\,\lambda1035
                     \else O\,{\sc vi}\,$\lambda1035$\fi}
\def  \OIVb         {\ifmmode {\rm O}\,{\sc iv]}\,\lambda1402
                     \else O\,{\sc iv]}\,$\lambda1402$\fi}
\def  \bOIIb       {\ifmmode {\rm [O}\,{\sc ii]}\,\lambda3727
                     \else [O\,{\sc ii]}\,$\lambda3727$\fi}
\def  \bOIb       {\ifmmode {\rm [O}\,{\sc i]}\,\lambda6300
                     \else [O\,{\sc i]}\,$\lambda6300$\fi}
\def  \OI       {\ifmmode {\rm [O}\,{\sc i]}\,\lambda1304
                     \else [O\,{\sc i]}\,$\lambda1304$\fi}
\def  \NII         {\ifmmode {\rm N}\,{\sc ii}\,\lambda1084
                     \else N\,{\sc ii}\,$\lambda1084$\fi}
\def  \NIII         {\ifmmode {\rm N}\,{\sc iii}\,\lambda990
                     \else N\,{\sc iii}\,$\lambda990$\fi}
\def  \NIIIb         {\ifmmode {\rm N}\,{\sc iii]}\,\lambda1750
                     \else N\,{\sc iii]}\,$\lambda1750$\fi}
\def  \NIVb         {\ifmmode {\rm N}\,{\sc iv]}\,\lambda1486
                     \else N\,{\sc iv]}\,$\lambda1486$\fi}
\def  \NV          {\ifmmode {\rm N}\,{\sc v}\,\lambda1240
                     \else N\,{\sc v}\,$\lambda1240$\fi}
\def  \MgII        {\ifmmode {\rm Mg}\,{\sc ii}\,\lambda2798
                       \else Mg\,{\sc ii}\,$\lambda2798$\fi}
\def  \CVI        {\ifmmode {\rm C}\,{\sc vi}\,368~eV
                       \else C\,{\sc vi}\,368~eV\fi}
\def  \SiIV         {\ifmmode {\rm Si}\,{\sc iv}\,\lambda1397
                     \else Si\,{\sc iv}\,$\lambda1397$\fi}
\def  \bFeXb       {\ifmmode {\rm [Fe}\,{\sc x]}\,\lambda6734
                       \else [Fe\,{\sc x]}\,$\lambda6734$\fi}
\def  \MgX        {\ifmmode {\rm Mg}\,{\sc x}\,\lambda615
                       \else Mg\,{\sc x}\,$\lambda615$\fi}
\def  \MgXI        {\ifmmode {\rm Mg}\,{\sc xi}\,1.34~keV
                       \else Mg\,{\sc xi}\,1.34~keV\fi}
\def  \MgXII      {\ifmmode {\rm Mg}\,{\sc xii}\,1.47~keV
                     \else Mg\,{\sc xii}\,1.47~keV\fi}
\def  \bNeVb      {\ifmmode {\rm [Ne}\,{\sc v]}\,\lambda3426
                     \else [Ne\,{\sc v]}\,$\lambda3426$\fi}
\def  \NeVIII      {\ifmmode {\rm Ne}\,{\sc viii}\,\lambda774
                     \else Ne\,{\sc viii}\,$\lambda774$\fi}
\def  \SiVIIa      {\ifmmode {\rm Si}\,{\sc vii}\,\lambda70
                     \else Si\,{\sc vii}\,$\lambda70$\fi}
\def  \NeVIIa      {\ifmmode {\rm Ne}\,{\sc vii}\,\lambda88
                     \else Ne\,{\sc vii}\,$\lambda88$\fi}
\def  \NeVIIIa      {\ifmmode {\rm Ne}\,{\sc viii}\,\lambda88
                     \else Ne\,{\sc viii}\,$\lambda88$\fi}
\def  \NeIX      {\ifmmode {\rm Ne}\,{\sc ix}\,915~eV
                     \else Ne\,{\sc ix}\,915~eV\fi}
\def  \NeX      {\ifmmode {\rm Ne}\,{\sc x}\,1.02~keV
                     \else Ne\,{\sc x}\,1.02~keV\fi}
\def  \SiXII        {\ifmmode {\rm Si}\,{\sc xii}\,\lambda506
                       \else Si\,{\sc xii}\,$\lambda506$\fi}
\def  \SiXIII      {\ifmmode {\rm Si}\,{\sc xiii}\,1.85~keV
                     \else Si\,{\sc xiii}\,1.85~keV\fi}
\def  \SiXIV      {\ifmmode {\rm Si}\,{\sc xiv}\,2.0~keV
                     \else Si\,{\sc xiv}\,2.0~keV\fi}
\def  \SXV      {\ifmmode {\rm S}\,{\sc xv}\,2.45~keV
                     \else S\,{\sc xv}\,2.45~keV\fi}
\def  \SXVI      {\ifmmode {\rm S}\,{\sc xvi}\,2.62~keV
                     \else S\,{\sc xvi}\,2.62~keV\fi}
\def  \ArXVII      {\ifmmode {\rm Ar}\,{\sc xvii}\,3.10~keV
                     \else Ar\,{\sc xvii}\,3.10~keV\fi}
\def  \ArXVIII      {\ifmmode {\rm Ar}\,{\sc xviii}\,3.30~keV
                     \else Ar\,{\sc xviii}\,3.30~keV\fi}
\def  \FeI_XVI      {\ifmmode {\rm Fe}\,{\sc 1-16}\,6.4~keV
                     \else Fe\,{\sc 1-16}\,6.4~keV\fi}
\def  \FeXVII_XXIII     {\ifmmode {\rm Fe}\,{\sc 17-23}\,6.5~keV
                     \else Fe\,{\sc 17-23}\,6.5~keV\fi}
\def  \FeXXV      {\ifmmode {\rm Fe}\,{\sc xxv}\,6.7~keV
                     \else Fe\,{\sc xxv}\,6.7~keV\fi}
\def  \FeXXVI      {\ifmmode {\rm Fe}\,{\sc xxvi}\,6.96~keV
                     \else Fe\,{\sc xxvi}\,6.96~keV\fi}
\def  \FeLa     {\ifmmode {\rm Fe}\,{\sc L}\,0.7-0.8~keV
                     \else Fe\,{\sc L}\,0.7-0.8~keV\fi}
\def  \FeLb     {\ifmmode {\rm Fe}\,{\sc L}\,1.03-1.15~keV
                     \else Fe\,{\sc L}\,1.03-1.15~keV\fi}
%
%     general symbols
%
\def  \L46         {$ L_{46} $}
\def  \Mo          {$ M_{\odot} $}     % Solar mass
\def  \Lo          {$ L_{\odot} $}     % Solar lumonosity
\def  \Fn          {$ F_{\nu} $}
\def  \Ln          {$ L_{\nu} $}
\def  \LEdd        {$ L_{\rm Edd} $}    % Eddington luminosity
\def  \MBH         {$ M_{\rm BH} $}     % black hole mass
\def  \m9          {$ m_9 $}
\def  \mdot        {$ \dot{m} $}        % accretion rate
\def  \Mdot        {$ \dot{M} $}        % accretion rate
\def  \Te          {$ T_{\rm e} $}      % electron temperature
\def  \Tex         {$ T_{\rm ex} $}     % excitation temperature
\def  \Teff        {$ T_{\rm eff} $}    % effective temperature
\def  \THIM        {$ T_{\rm HIM} $}    % Temp. hot inter. medium
\def  \TC          {$ T_{\rm C} $}      % Compton (or color) temperature
\def  \Ne          {$ N_{\rm e} $}      % electron density
\def  \nh          {$ n_{\rm H} $}      % hydrogen density
\def  \N10         {$ N_{10} $}
\def  \Ncol        {$ N_{\rm col} $}    % column density
\def  \rin         {$ r_{\rm in} $}
\def  \rout        {$ r_{\rm out} $}
\def  \rav         {$ r_{\rm av} $}
\def  \jc          {$ j_c(r) $}
\def  \Rc          {$ R_c(r) $}
\def  \Ac          {$ A_c(r) $}
\def  \nc          {$ n_c(r) $}
\def  \Cr          {$ C(r) $}
\def  \Cf          {$ \rm C_f $}     % covering fraction
\def  \El          {$ E_l(r) $}
\def  \el          {$ \epsilon_l(r) $}
\def  \PSI         {$ \Psi (t) $}
\def  \me          {$ m_{\rm e} $}            % electron mass
\def  \mp          {$ m_{\rm p} $}            % proton mass
\def  \rBLR        {$ r_{\rm BLR} $}          % BLR radius
\def  \rNLR        {$ r_{\rm NLR} $}          % NLR radius
\def  \aox         {$ \alpha_{ox} $}          %
\def  \Ux          {$ \rm U_{\rm x} $}  % X-ray ionization parameter
%
% units
%
\def  \kms         {\hbox{km s$^{-1}$}}          % kilometers per sec
\def  \ergs        {\hbox{erg s$^{-1}$}}              % erg/sec
\def  \ergsHz      {\hbox{erg s$^{-1}$ Hz$^{-1}$}}   %  erg/sec/Hz
\def  \cc          {\hbox{cm$^{-3}$}}
\def  \cmii        {\hbox{cm$^{-2}$}}
\def  \cms         {\hbox{cm s$^{-1}$}}      % cm / sec
\def  \mic         {$\mu$m}
\def  \vs          {{\it vs.} }
\def  \etal        {{\it et al.}}
%
% Ian's maths macros
%
\newbox\grsign \setbox\grsign=\hbox{$>$} \newdimen\grdimen \grdimen=\ht\grsign
\newbox\simlessbox \newbox\simgreatbox \newbox\simpropbox
\setbox\simgreatbox=\hbox{\raise.5ex\hbox{$>$}\llap
     {\lower.5ex\hbox{$\sim$}}}\ht1=\grdimen\dp1=0pt
\setbox\simlessbox=\hbox{\raise.5ex\hbox{$<$}\llap
     {\lower.5ex\hbox{$\sim$}}}\ht2=\grdimen\dp2=0pt
\setbox\simpropbox=\hbox{\raise.5ex\hbox{$\propto$}\llap
     {\lower.5ex\hbox{$\sim$}}}\ht2=\grdimen\dp2=0pt
\def\simgreat{\mathrel{\copy\simgreatbox}}
\def\simless{\mathrel{\copy\simlessbox}}
\def\simprop{\mathrel{\copy\simpropbox}}
\def \GINGA  {{\it GINGA}}
\def \Ginga  {{\it Ginga}}
\def \ASCA   {{\it  ASCA}}
\def \asca   {{\it  ASCA}}
\def \BBXRT  {{\it BBXRT}}
\def \ROSAT  {{\it ROSAT}}
\def \XTE    {{\it XTE}}
\def \XLi    {XL$_1$}
\def \XLii   {XL$_2$}
\def \XLiii  {XL$_3$}  
\def \XMi    {XM$_1$}
\def \XMii   {XM$_2$}
\def \XMiii  {XM$_3$}
%
%
% Journals
%
\def \aap     { {\it Astr. Ap.}, }
\def \aaps    { {\it Astr. Ap. Suppl.}, }
\def \aj      { {\it A. J.}, }
\def \apj      { {\it Ap. J.}, }
\def \apjl     { {\it Ap. J. (Letters)}, }
\def \apjs    { {\it Ap. J. Suppl.}, }
\def \iau      { {\it International Astronomical Union} }
\def \mnras   { {\it M.N.R.A.S.}, }
\def \pasp    { {\it Pub.A.S.P.}, }
\def \BAAS  { {\it Bulletin of the American Astronomical Society}}
\def \Icarus  { {\it Icarus}}
\def \JMolSpec  { {\it J. Molec. Spectrosc.}}
\def \JQSRT  { {\it J. Quant. Spectrosc. Rad. Trans.}}
\def \Science  { {\it Science}}
\def \JGR  { {\it J. Geophys. Res.}}
\def \PhDthesis  { {\it Ph. D. thesis}}
\def \JOptSocAm  { {\it J. Opt. Soc. Am.}}

\begin{abstract}
Previous X-ray and ultraviolet spectroscopy suggested
 that the Fe/O abundance ratio in  
 NGC~1068 may be abnormally high. We have tested this suggestion by 
measuring and modeling the {\it ASCA} spectrum of NGC 1068. We have measured 
some 15 X-ray lines, to an accuracy better than a factor 2,  
and modeled the continuum in two different ways. The first assumes 
that the hard X-ray continuum is the reflection of the nuclear 
source by two, extended, photoionized 
gas components; a warm (T$\sim 1.5 \times 10^5$ K) gas and a hot
(T$\sim 3 \times 10^6$ K) gas. All the observed emission lines are produced
in this gas and there is an additional, extended 0.6--3 keV pure continuum
component. The model is similar to the one proposed by Marshall \etal\ (1993).
The second model is a combination of a hard reflected continuum
with a soft thermal plasma component.
The calculations show that the emission lines in the photoionized gas model
are in very good agreement with the observed ones assuming solar metallicity 
for all elements except for iron, which is more than twice solar, and oxygen,
which is less than 0.25 solar. Models with solar oxygen are possible if the
0.5--1 keV continuum is weaker but they do not explain the magnesium and
silicon lines. The thermal model fit requires extremely
low metallicity (0.04 solar) for all elements. We discuss these findings 
and compare them with the {\it ASCA} spectra  of recently observed 
 starburst galaxies. 
We argue that the apparent low metallicity of starburst galaxies, as well
as of the extended nuclear source in NGC~1068, are inconsistent with galaxy
chemical evolution. The explanation for this apparent 
anomaly is still unknown and may 
involve  non-thermal continuum mechanisms and, in some cases, 
depletion onto grains. Given  
the strong H-like and He-like lines,
  as well as the prominent Fe-L emission features, the 
origin of the soft X-ray lines in this source 
is more likely photoionized gas.
We  compare 
our model with the recent Iwasawa \etal\ (1997) paper. We also show that   
  fluorescence lines of low-Z element lines in AGN is likely to be larger
than previously assumed.
\end{abstract}

\section{INTRODUCTION}
NGC~1068 is perhaps the best studied Seyfert 2 galaxy, with X-ray observations
by {\it EXOSAT}, {\it GINGA}, {\it ROSAT}, {\it BBXRT} and {\it  ASCA} 
(for references and general description
see Marshall \etal; 1993). The most striking X-ray feature is the strong
(equivalent width, EW$=\simeq 3$~keV) \Ka\ complex that is split into three
components corresponding to ``neutral'' (6.4--6.5 keV), He-like (6.7 keV)
 and H-like (6.96 keV) iron lines. The large EW is probably the result of the
obscured central continuum and the directly viewed line producing gas (Krolik
\& Kallman; 1988). This iron line complex was first resolved in 
{\it BBXRT} data (Marshall \etal\ 1993) and was later studied by 
Ueno \etal\ (1994) and Iwasawa, Fabian and Matt (1997). 

Marshall \etal\ (1993) suggested a two component 
model to explain  the X-ray emitting gas in NGC~1068. 
 The first  is a ``warm'' component with a typical temperature
of 2$\times 10^5$ K, and the second a more highly ionized ``hot'' 
component, with
 T$\sim 4 \times 10^6$ K. Both components are ionized
by the central X-ray source and co-exist, with pressure equilibrium, throughout
the nucleus. Both components reflect the optical-ultraviolet-X-ray continuum, 
and the broad Balmer lines, and are thus the  ``electron scattering mirror'' 
in this source. That  analysis 
showed that the iron abundance, as calculated from the EW of the 6.4 keV line,
is high, 2--3 times solar. The absence of a detectable \OVIII\ line has been
interpreted, within the framework of the model, as indication for
a very small O/H.
The implication is that the O/Fe abundance ratio is very
small,   an order of magnitude below  solar.  Ueno \etal\ 
(1994)  measured a very weak \OVIII\ line  and confirmed the
{\it BBXRT}  measurements of the \Ka\ lines. These authors took a 
different approach and modeled the observed \Ka\ lines, as 
well as several softer lines, by a 
 thermal plasma. Their underlying assumption is that  collisionally
ionized gas is producing the observed emission lines. However, such 
models require  abnormally low metallicity  since the predicted
soft X-ray lines are much stronger than those observed.

Marshall 
\etal\ have also noted the extreme weakness of the \OIIIb\ ultraviolet line
in this galaxy and suggested that the oxygen abundance anomaly  extends
 to the cooler gas, in the narrow line
region (NLR). The ultraviolet spectrum has been investigated, in more detail,
 by Netzer (1997), reaching
a similar  conclusion about the O/C and O/N abundance ratio.   

We have undertaken a more detailed study of the spectrum of NGC~1068, aiming
at  better constraints on the soft X-ray lines and the chemical
composition of the X-ray emitting gas. We have used the recent {\it {\it ASCA}} data
set, as described in \S2, and  measured many soft X-ray
lines. We have modeled the gas in various ways, as discussed in \S3 and
the findings are  discussed in \S4.

\section{DATA SELECTION AND ANALYSIS}

NGC~1068 was observed by {\it {\it ASCA}} on 1993 July 25
with an on-source time $\sim 39$ ksec. 
The \asca\ satellite and instruments are
described in Tanaka, Inoue \& Holt (1994).  
In summary, four co-aligned, grazing-incidence,
foil-mirror telescopes 
direct X-rays simultaneously 
onto four focal-plane instruments. There
are two CCD detectors -- the Solid-state Imaging Spectrometers (SIS)
 and two gas--scintillation proportional--counters --
the Gas Imaging Spectrometers (GIS).

NGC~1068 was observed with the SISs in 4-CCD mode, with data
accumulated in both 'FAINT' and 'BRIGHT' telemetry modes.  As 
NGC~1068 has a relatively low count rate in the SIS,
the superior resolution available in
'FAINT' mode could not be utilized. Therefore the FAINT and BRIGHT mode
data were combined for the analysis presented here.

We extracted the ``raw'' event files from the {\it ASCA} 
archive and used these as a starting point for our analysis.
These files were created from the original 
telemetry data and have been corrected to produce linearized detector
coordinates, gain corrected pulse--height values and sky
co-ordinates determined from the spacecraft aspect solution. 
We applied data selection and cleaning algorithms using
FTOOLS/XSELECT v3.5. 
Data were rejected by removing 'hot' and 'flickering' pixels in the SISs;
removing data accumulated during  passages through the South Atlantic 
Anomaly; imposing a minimum geomagnetic rigidity of 6 GeV/$c$; 
removing data accumulated when the angle from the Earth's limb was less
than 20$^{\circ}$ during orbit-day and less than  $10^{\circ}$ (SIS) or 
$5^{\circ}$ (GIS) during orbit-night;
restricting SIS data to event 'GRADES' 0,2,3 and 4 and rejecting data taken 
within 200 seconds after crossing the day/night terminator.
We combined the {\tt LOW}, {\tt MEDIUM} and {\tt
HIGH} bit-rate data.

Application of these screening criteria gave a mean effective exposure time
of $\sim$39 ks in the GIS instruments and $\sim 10$ ks in the SIS 
instruments. The SIS exposures were significantly lower because 
necessary differences in the selection criteria for the two instruments, 
SIS data in 4--CCD mode are particularly prone to 
problems with telemetry saturation during periods of LOW bit-rate, these 
caused data dropouts and hence a reduction in useful exposure time. 

Images were extracted from the screened and cleaned data
from all instruments, and region descriptors defined for  the extraction of
light curves and spectra. For the two SIS instruments, we used a
3~arcmin circle centered on NGC~1068  with the background taken 
towards the edge of the same CCD chip.
For the two GIS instruments, we used a circular extraction cell of
5~arcmin radius centered on NGC~1068  with the background taken in
a nearby source-free region. 

The \ASCA\ light curves reveal  no significant flux variability 
across the observation, and thus we consider only the mean spectrum
in this paper.

Data from both pairs of SIS and GIS instruments were analyzed together,
but with the normalization of each  data set allowed
to vary relative to the others (since there are small
discrepancies in the absolute flux calibrations of the detectors).
We used the SIS response matrices released November 1994, and the GIS 
response matrices released March 1995.

\section{MEASUREMENT AND ANALYSIS}
\subsection{Empirical Fits}
Inspection of the \ASCA\ spectrum clearly shows a hard continuum source,
 strong \Ka\ lines,
 several lower energy emission features and a soft 0.5--3 keV continuum
component (Marshall \etal; 1993, Ueno \etal; 1994).
 As argued by Wilson \etal\ (1992), 
 the soft 0.5--3keV  emission is from an extended source,
which is contributing at least half the flux at those energies. 
The likely explanation is an extra-nuclear 
starburst region.  This would imply that the 
  soft  X-ray central source is weaker  than the one assumed in  Marshall \etal\
(see an extensive
discussion by Pier \etal, 1996).

Our approach in this work is to first measure the soft X-ray lines in
a way which is independent of any  model and then
to compare those measurements with the prediction of several specific 
 models. 
The first step is to  fit the 
 0.6--10 keV continuum and all significant emission features that
 do not correspond to known detector features.
The  procedure assumes that the observed continuum 
 can be approximated  by  two
smooth functions, such as two power-laws or an absorbed power-law and a thermal
continuum. There are several combinations that produce reasonable fits and we do
not attach great significance to the chosen functions since they do not 
represent physical entities. We accept any combination that fits well the
overall underlying spectrum. The soft component was allowed  to vary 
freely in the fit, while the hard component was allowed to range between 
1.5 and 1.7 in slope.
Fig. 1 shows an example where the high energies are
  fitted a power-law with absorbed
soft X-ray part and the low energies are fitted by a second,
 unabsorbed power-law. In
this  example, the hard component  photon $\Gamma$ slope is 1.6 
($F_{N(E)} \propto E^{-\Gamma}$) and
the soft component slope is $\Gamma=3.4$. Other combinations give 
equally good fits where in all cases the hard X-ray photon 
slope  is between 1.5 and 1.7. 
\begin{figure}
\epsfxsize\hsize
\epsfbox{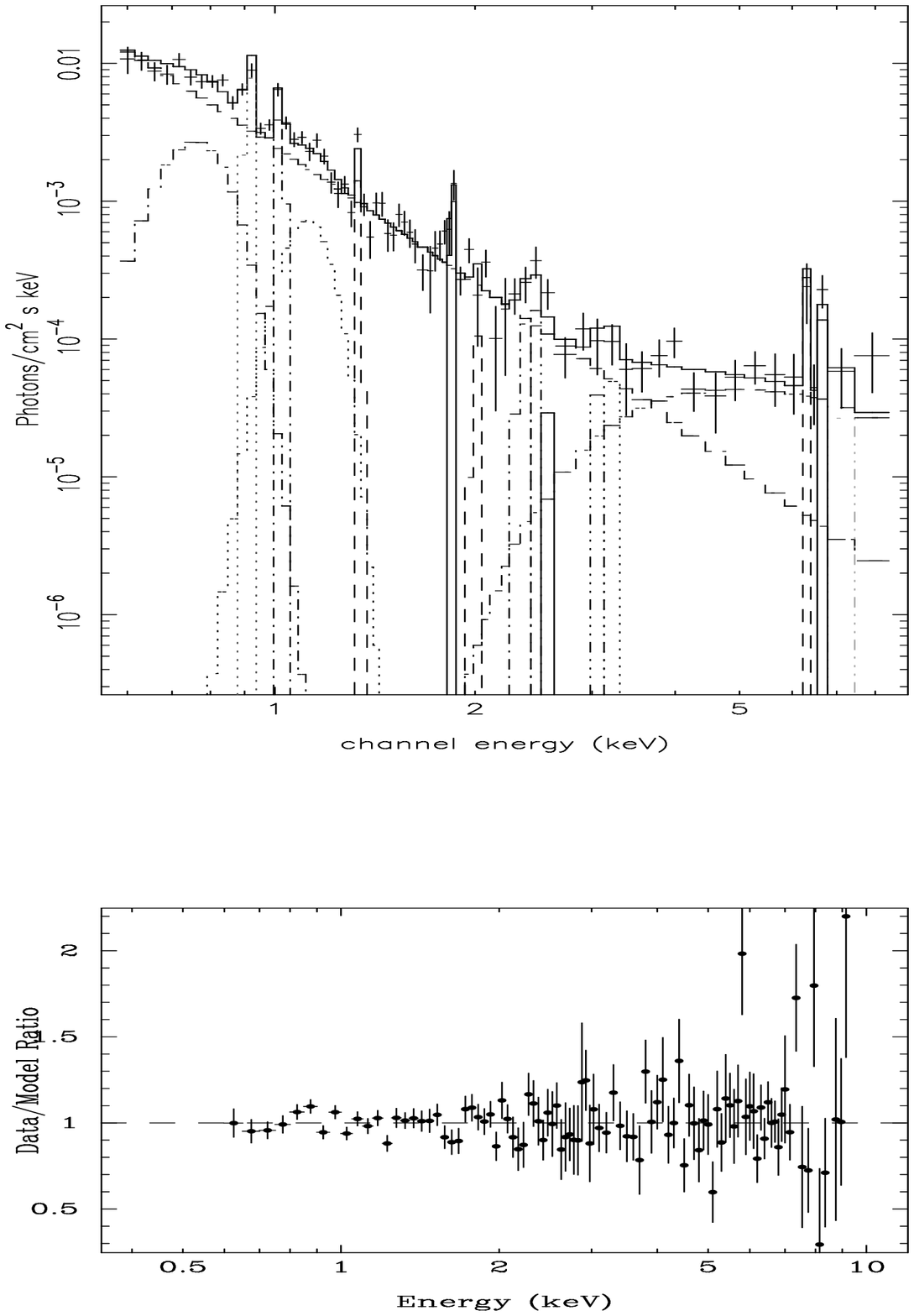}
\caption{
 \ASCA\ X-ray spectrum and fitted model for
 NGC~1068. The various model components are shown at the top and the
residuals (data/model) at the bottom.}
\end{figure}

 Next, we have 
added a large number of narrow gaussian lines
 as free parameters. These include the strongest
H-like and He-like lines of all elements in ION  and about 20
Fe-L lines. A large number of those are crowded in a small energy range, 
especially over 
the ranges of 0.7--0.9 and 1--1.2 keV, which contain 
a plethora of iron-L lines. The \ASCA\ resolution at those energies
is not sufficient to model individual lines 
 and we have fitted these regions  by  
broad gaussians, where the width and centroid energies were chosen 
based on the spread on the predicted shape of the line blends. 
Some line pairs (e.g. the 
H-like and He-like magnesium lines) are also close in energy and while we
attempt to fit them individually, in the error analysis (see below) we consider 
only the combined strength. The total number 
of broad and narrow gaussians, including the 3 iron \Ka\ lines,
  is 15. Fig. 1 shows one of the best fits
and Table 1 lists the line intensities  corresponding to it.

\begin{table}[t]
%\begin{minipage}{19cm}
  \caption{Observed and calculated X-ray line intensities}
 \renewcommand{\arraystretch}{0.75}
\begin{tabular}{lcccc}
\hline
\hline
Line       & Observed flux\footnote{One flux unit corresponds to $7.4 \times 10^{-13}$ ergs cm$^{-2}$s$^{-1}$}
  & Model 1\footnote{Low oxygen, break at 1 keV}& Model 2\footnote{Solar oxygen, break at 0.5 keV}
 & 90\% confidence interval \\
\hline
\OVII      &         & 0.47 & 0.54 & \\
\OVIII     & $<0.4$  & 0.30\footnote{The calculated EW,
   relative to the observed continuum, is 23 eV.}& 0.26 &  $<0.4$   \\
\FeLa      &0.98  & 0.42& 0.13 & \\
\NeIX  &0.61 &0.36& 0.09   \\
0.72 - 0.92 keV  & 1.11 &0.78& 0.22 &1.08 - 2.1  \\
\NeX       &0.32  &0.08& 0.05  \\
\FeLb      & 0.35  &  0.18& 0.18 \\
total 1.0 - 1.2 keV  &  0.67 &0.26 & 0.23 &0.47 - 0.86  \\
\MgXI      & 0.09 & 0.08 & 0.04  \\
\MgXII     & 0.0 &  0.03 & 0.03 & \\
total 1.34 - 1.47 keV&0.12 &0.11 &0.07  & 0.07 - 0.17 \\
\SiXIII    & 0.16 &0.07 & 0.02&   \\
\SiXIV     & 0.03 & 0.05& 0.05&  \\
total 1.85 - 2.01 keV &0.19 &0.12 & 0.07 & 0.11 - 0.28 \\
S\,{\sc i}-S\,{\sc x} 2.35 keV  & 0.07 &0.02 & 0.02 &  \\
\SXV       & 0.07 &  0.02& 0.01&  \\
\SXVI      & 0.02 & 0.04 & 0.04& \\
total 2.35 - 2.62 keV &0.16 &0.08 & 0.07 &0.02 - 0.32\\
\ArXVII    &0.04  &  0.01 & 0.01&  \\
\ArXVIII   & 0.06 &  0.02 & 0.02 & \\
total 3.1 - 3.31 keV &0.10 & 0.03 & 0.03 &  0.01 - 0.19 \\
Fe\,{\sc i}-Fe\,{\sc xvi} 6.4 keV & & 0.5 &  0.5 & \\
Fe\,{\sc xvii}-Fe\,{\sc xxiii} 6.5 keV &  &  0.36 & 0.15 & \\
Fe\,{\sc i}-Fe\,{\sc xxiii} 6.4--6.5 keV &0.86  &0.86  & 0.65 & 0.63 - 1.14 \\\FeXXV     & 0.54\footnote{The  EW of the 6.7 keV
            line is 1.05 keV} &  0.50 & 0.5 & 0.29-0.83 \\
\FeXXVI    & 0.29 &  0.30 & 0.29 & 0.09 - 0.52 \\
\hline
\end{tabular}
%\end{minipage}
\end{table}

Error estimates are carried out separately for lines and continuum.
The continuum uncertainties are not very important since the acceptable
range of slopes  is rather small (less than 0.2 dex). As for the
lines, we have estimated the 90\% confidence interval for each line. 
The $\chi^2$ step which corresponds to the 90\% error range for any 
particular line, depends upon the number of free parameters which 
are interdependent with that line normalization. This in turn 
depends on the relative proximity of other lines, and where 
the line falls in the spectrum.  
The line and line-blend estimates are listed in Table 1. 

Next we introduce various physical models that represent some combinations
of photoionized and collisionally ionized plasmas.

\subsection{Models Involving Photoionized Gas}
We have investigated the possibility that the observed soft X-ray lines
are due to photoionized gas in the nucleus of NGC~1068. We have tried
this idea in two different ways: by looking at line intensities in a specific,
two component photoionization model, and by scanning a grid of models in
a search for the best combination of ionization parameter and column density that fit
the {\it ASCA} spectrum.

The  specific photoionization model assumes that all emission lines 
originate in two distinct photoionized gas component, a highly ionized component (hereafter the ``hot''component) and a moderate ionization component
(hereafter the  ``warm'' component).
 Regarding the continuum,
the assumption is that there is    a ``hard'' 
(1--10 keV), obscured nuclear part, typical of Seyfert galaxies, 
and a ``soft'' ($0.5<E<$3 keV) extended part which is of unknown origin
and plays no part in exciting the nuclear gas.
Thus, the observed continuum is made out of three components: 1. The reflection
of the nuclear continuum by the hot photoionized gas. 2. The reflection
of the nuclear continuum by the ``warm'' photoionized gas, and 3. The extended,
directly observed continuum. The soft component contributes 
a negligible amount above 
5 keV. Thus, the  normalization  is  such 
 that the two reflected components (in about equal amounts, see below) 
fit the observed {\it ASCA} and {\it GINGA} 
high energy continuum and that the combination 
of all three gives a good representation of the overall continuum
energy distribution.  
  
The spectral energy distribution (SED) of the 
central continuum  is similar to the one used
 by Marshall \etal\ (1993), except for the soft X-ray part. It
is made of an infrared-optical-UV broken power-law that 
  fits the overall energy budget of NGC 1068 and an X-ray  power-law of
photon slope $\Gamma=1.5$ that fits the 1--50 keV continuum. The
X-ray  is added smoothly to the UV part, below 1 keV.
 The resulting \aox, without taking into account the soft extended
continuum, is 1.6. 
 The  nuclear continuum produces 
about 15\% of the total flux at 1 keV and the rest is assumed to be
extended (i.e. the 15\% is the fraction of the reflected nuclear
source out of the total, reflected plus directly observed extended flux). 

We have also tried to extend the nuclear component, with the same slope,
to lower energies (0.5 keV). 
 As shown below, this helps the oxygen abundance anomaly but introduces
other difficulties. The exact optical-UV continuum properties are not very important
for the purpose of the present discussion but are influencing the observed
ultraviolet lines (see Netzer, 1997).

We have used the photoionization code ION (Netzer 1996 and references
 therein) to 
 model  the X-ray ionized  gas in NGC~1068. The model inputs are the assumed
central continuum, the density and column density, the X-ray 
ionization parameter (i.e. as defined by the 0.1--10 keV photon flux, see 
Netzer 1996) and the abundances. We have tested two
possible compositions. The first is motivated by the suspected
unusual oxygen and iron abundances and assumes the
following dust-free composition:\\
H:He:C:N:O:Ne:Mg:Si:S:Ar:Fe=
 $10^{4}$:10$^3$:3.4:1.1:1.5:1:0.35:0.35:0.16:0.07:1.2, 
i.e. solar except for iron, which is three times solar and oxygen, which is
0.2 solar.  As shown
below, this is in good agreement with the observations. The second is the same
except that O/H is solar (8$\times 10^{-4}$).
 
The photoionization model which is compared with our line measurements
is  similar to the one presented by
 Marshall \etal\ (1993). The warm component column density is
 5$\times 10^{22}$
\cmii\ with an illuminated face density of N$_{\rm H}=700$ \cc\ and the hot component
column is 1.9$\times 10^{22}$ \cmii,
 with illuminated-face density of 3100 \cc. In both models
${\rm N_H \propto R^{-1.5}}$, which is required to explain the large spatial
extent of the gas (the optical ``mirror'').
 The illuminated face X-ray ionization
parameters are 3.5 and 310, for the warm and hot components 
respectively (for the physical dimensions see Marshall \etal\ 1993).
 We have not introduced any expansion motion  but assumed local
microturbulences that correspond to the  local sound speed.
 This has some minor effects on the observed
spectrum  since in this reflection-only
geometry, continuum fluorescence can influence some observed line intensities
(Krolik and Kriss 1995;  Netzer 1996).

Having defined the gas density, location and abundance, we
 have calculated the expected X-ray spectrum of the two components.
The normalization of the reflected and emitted spectrum of the hot and warm
components (i.e. the flux unit in Table 1) 
is such that the calculated \Ka\ lines match the
observed values (Table 1, Model 1). The two separate components, and the
resulting composite spectrum, are shown in Fig. 2. Also shown is a comparison
of the model (reflected nuclear continuum) with the observations (reflected plus
the directly observed extended continuum).
\begin{figure} 
\epsfxsize\hsize 
\epsfbox{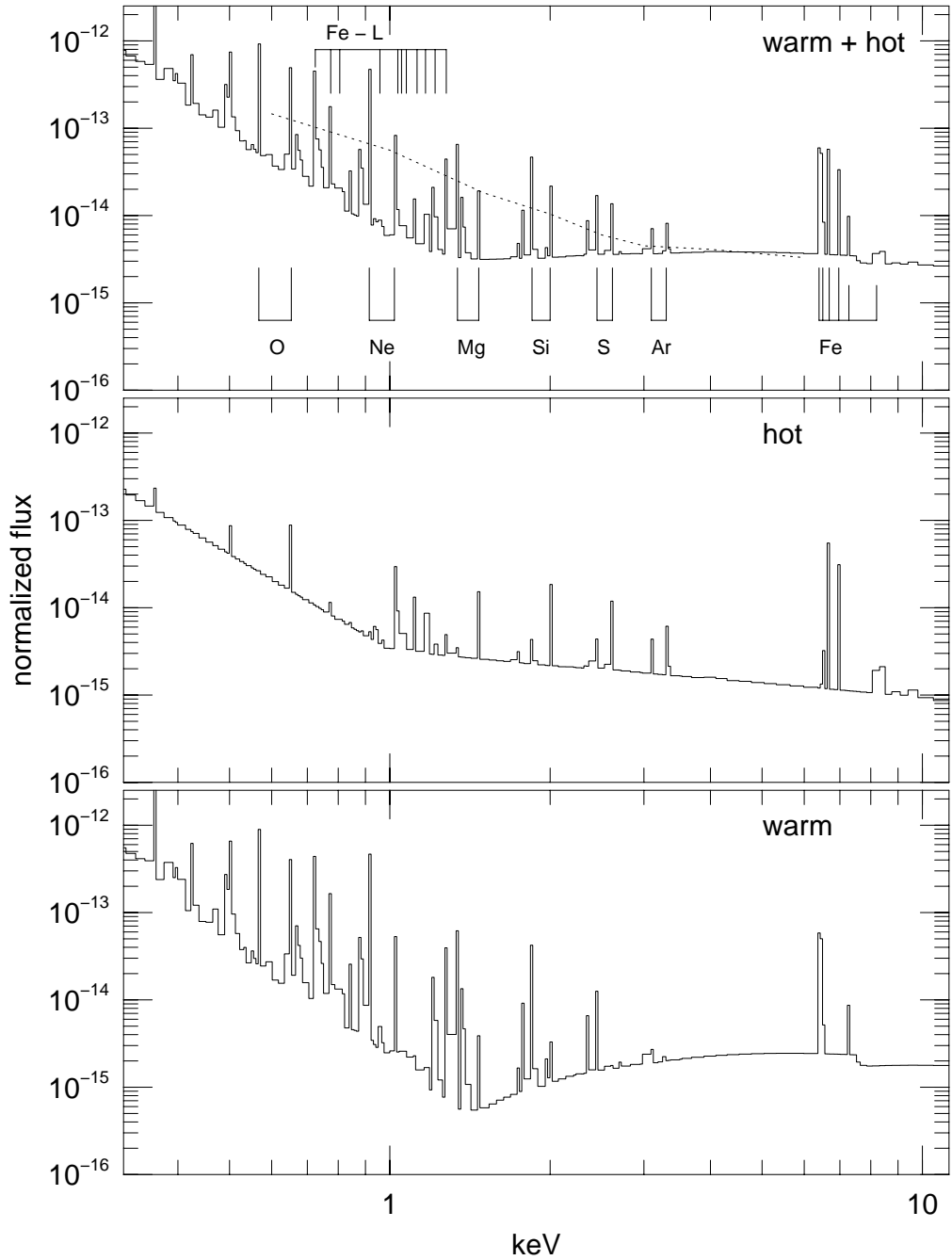} 
\caption{ 
 Calculated warm (bottom), hot (middle)
  and combined (top)  photoionized gas spectra for NGC~1068 assuming
low O/H and continuum break at 1 keV.
 The strongest emission lines are marked and the observed continuum is shown
at the top in dashed line. The difference between the observed and calculated
continua is attributed to the extended component.
 }
\end{figure} 

As evident in Fig. 2, the warm gas component  exhibits a rich soft
X-ray spectrum, composed of numerous lines and edges, and  strong ``neutral''
\Ka\ lines at 6.4--6.5 keV. The strongest soft X-ray  features
are \NeIX, \MgXI, \SiXIII\ and several Fe-L lines. 
There is  a noticeable broad absorption feature, centered 
around 1.5 keV, which is
due to the combined opacity of neon and iron. Oxygen opacity is not important
because of the low assumed oxygen abundance. The same is true for the strength
of the \OVIII\ line which, with the assumed composition, is consistent with the
measured upper limit.  The volume averaged gas temperature in this component
is about 1.1$\times 10^5$ K and the highest (illuminated face) temperature
is 1.5$\times 10^5$ K. 
The hot component spectrum is of much higher ionization and temperature 
(1.5--3.9$\times 10^6$ K). 
This gas produces strong He-like and H-like iron lines and the strongest
soft X-ray lines are \NeX, \MgXII, \SiXIV, \SXVI\ and a few Fe-L lines.
 
Table 1 gives the  calculated line intensities,
normalized as explained,  and compares them with the 
measurements. It shows that the assumed continuum shape, ionization
parameter and metalicity, produce a good fit to the observed spectrum.

We have searched for ways to eliminate the need for the abnormally low O/H and 
found that it is strongly dependent on the 0.5--1 keV incident flux.
We have therefore tried a fit assuming a nuclear source with a $\Gamma=1.5$ 
slope 
covering the entire 0.5--50 keV range, i.e. much weaker than the previously
guessed continuum below 1 keV. In this case, oxygen is less ionized
and the \OVIII\ equivalent width is consistent with the observed upper
limit with solar O/H. However, the model suffers from several inconsistencies.
First, the calculated 0.7--0.8 keV Fe-L line fluxes are 
well below their observed intensity, 
despite the large assumed Fe/H. Second, some other lines, most notably
\NeIX, \MgXI\ and 
\SiXIII\, are much below their observed flux. Thus, in this case, the
abundances of neon, magnesium and silicon are all problematic.
 As seen in Table 1, the overall fit in this case is much less
satisfactory.

The second approach is to calculate a grid of hot and warm 
photoionization models,
 to combine them with an additional,
 soft continuum component, and to use a minimization
technique to search for the  combination that fit best the observed spectrum.
This three component fit is chosen from a grid
of models (presented as {\it atables} in the fitting routine {\it XSPEC})
 covering a range of column density and X-ray ionization
parameter, and assuming a single-slope 0.5--50 keV continuum.
 The  soft soft excess is modeled as thermal bremsstrahlung
emission (to avoid emission lines, see below)
 with temperature as the only free parameter. The photoionized gas composition
 is identical to the one shown above. Using these
components, we find a reasonable good fit with gas
 kT=0.44 keV (T=5.1$\times 10^6\,$K) plasma, column
densities of 10$^{22.4}$ \cmii\ (warm) and 10$^{22.6}$ \cmii\ (hot) and X-ray ionization parameters of 1.3 (warm)
and  95 (hot). These ionization parameters are measured at the illuminated
face and are thus slightly above 
  the volume averaged ionization
parameters of the   ${\rm N_H \propto R^{-1.5}}$  models.
 The reduced $\chi ^2$ is 1.37 for 509 degrees of freedom.
We do not show this fit since it is similar in quality, and major features,
to the one shown in Fig. 1. We also note that allowing little freedom in
the assumed composition can significantly improve the fit. We avoid this
additional complication since, with the limited {\it ASCA} capability,
 it adds nothing to the understanding of the source. 

\subsection{Models Involving Collisionally Ionized Gas}
Our next model is meant to test the hypothesis that most of the {\it line
and continuum} soft flux is due to the extended, starburst region. The purpose
 is to test the required composition and overall 0.6--3 keV continuum shape 
and not to model individual lines. 

We have modeled the soft spectrum by a hot plasma ``meka'' model.
The meka model describes an 
emission spectrum from hot diffuse gas based on the model calculations
of Mewe \& Gronenschild (1985), 
Mewe, Lemen,  \& van den Oord (1986), 
and Kaastra (1992). The model inputs are plasma temperature 
 hydrogen density and composition. Meka includes line
emission from C, N, O, Ne, Na, Mg, Al, Si, S, Ar, Ca, Fe and Ni. 
We assumed a gas density of 10$^3$ \cc\ and allowed the plasma 
temperature and composition to vary. 
We added a $\Gamma=1.5$ continuum and a broad Gaussian 6.5 keV
line to fit the hard X-ray  emission and to enable a 
reasonable parameterization of the overall X-ray spectrum and hence 
achieve a meaningful $\chi ^2$ 
minimization. The best fit is obtained with kT=0.59 keV and metallicity of 
0.043 solar. We have  measured the line emission between 0.6  and 
1 keV, which represents most of the line flux in our best fit model.
 The measured equivalent 
width (EW) of the line blend  
relative to the 0.59 keV continuum, is about 400 eV, i.e. similar
to the total emission line EW measured by our two photoionization models. 
This verifies that models with higher metallicity gave unacceptable fits
since they produce too much line emission. This is of great importance to
NGC 1068, and to starbursts in general, as discussed in the following section.

\section{DISCUSSION}
Our data analysis and model fitting  show
clear evidence for several strong soft X-ray lines in the spectrum of NGC~1068. We have measured those with typical
 uncertainty of a factor two. The  EWs  are large, compared
with those predicted for Seyfert 1 galaxies (Netzer 1996).
 This indicates either 
collisionally ionized plasma or else photoionized gas seen against a reflected
continuum. Below we examine the consequence of both as well as the 
new model by Iwasawa \etal\ (1997).

\subsection{Photoionization models}

The two component photoionized gas model, with small O/H, large Fe/H and continuum
break at 1 keV, 
 nicely reproduces the observed line intensities in NGC 1068, given the  
 uncertainties. The  Marshall \etal\ (1993) result of large Fe/H  
 is  confirmed by our fitting of the \Ka\ complex as well as by the 
measured intensity of the Fe-L lines. Given this continuum,
the observed upper limit
on the \OVIII\ line intensity suggests that  O/H is 
smaller than about 0.25 solar, again in agreement with  Marshall
 \etal\ 
The composition of all other metals is 
within a  factor two of solar.

Reducing the nuclear flux by moving the break to 0.5 keV, considerably
change the conclusion about O/H. This is now consistent with solar but other 
lines are in bad agreement with the observations. In particular, we could
not find a satisfactory fit with solar Ne/H, Mg/H and Si/H. The key future observation is the
\OVII\ line which is predicted to be strong in both cases (see Table 1).
For example, the line ratios \OVII/\NeIX\ and \OVIII/\NeIX\ could be used to confront
the two possibilities. 
AXAF has got the capability to observe this ratio. 

The unusual gas composition required by our  photoionized 
gas models is a severe theoretical
 problem, since large enhancement of all metals, including oxygen,
is expected in central 
regions of evolved galaxies.
 It is possible that the poor calibration of 
the \ASCA\ detectors below 0.6 keV, and the low signal-to-noise of 
{\it BBXRT}, 
are the origins of the discrepancy. Alternatively,  the models shown here 
 do not represent the physical conditions in  
 NGC~1068. However, the presence of a rich spectrum of other metal
lines, all consistent with each other, hint to some anomaly in the O/Ne,
O/Mg, O/Si and O/S ratios, and in particular in O/Fe. We note again 
 the unusually weak 
 \OIIIb\ line in this galaxy, compared with the intensity of semi-forbidden
lines of nitrogen and carbon. As explained in 
Netzer (1997), this is not observed in four 
other Seyfert galaxies, with measurable ultraviolet narrow lines, and is
consistent with low O/N and O/C. Millimeter observations of this galaxy, also
hint to an unusual composition (Sterenberg, Genzel, and Tacconi, 1995).

The warm photoionized gas component in NGC 1068 is similar in column density
and ionization parameter to many ``warm absorber'' systems observed in
Seyfert 1 galaxies (see George \etal\ 1997 for references and review). 
It is interesting to examine the spectrum of the present warm component when
observed 
against a bare central continuum, since the Seyfert 1 systems are normally
assumed to originate in gas which is much closer to the central source.
 We have examined this possibility and 
found some differences as well as various dynamical implications.
The issue has been addressed by Krolik and Kriss
(1995) and Netzer (1996) and is beyond the scope of the present paper.  

Our photoionization model requires a third, pure continuum component to explain
the 0.5--3 keV excess. 
As discussed below the origin of this component is unclear. 

\subsection{Collisionally Ionized Plasma}

We have attempted to fit the soft 
X-ray flux by hot plasma models with variable composition. Our single
temperature solution requires extremely low (0.04 solar) metallicity.
A similar difficulty was encountered by  Ueno \etal\ (1994), who have attempted
a multiple component fit to the spectrum. Their best solution is 
composed of two hot plasma 
components one of which has 0.03 solar abundance. The origin 
of this problem is the large EW of lines associated with such a hot plasma.
We do not consider  such models adequate for NGC 1068, or other evolved
regions in galactic nuclei. No optical observations support this and
we know of no theoretical model to support this. We note that Pier
\etal\ (1996) suggested that much of the soft excess in NGC 1068 
is due to bremsstrahlung emission from the optical ``mirror''.
 This radiation is specifically
 included in our calculation and, as seen in Fig. 2, 
cannot explain the soft excess. 

The apparent composition anomaly  is  common to other starburst galaxies. 
As shown by Serlemitsos, Ptak and Yaqoob (1995), several other sources
show soft X-ray lines which, when fitted with multi-component hot plasma
models indicate very low metallicity. Interesting examples
are shown by Ptak \etal\ (1997) who studied M82 and NGC 253. Their deduced
metallicities are 10--30\% solar. We have remeasured the M82 {\it ASCA} data
used by Ptak \etal\ and found several soft X-ray lines with typical EWs in
the range of 50--100 eV, i.e. similar to those observed in NGC 1068. 
In the context of 
hot plasma models, the lines indicate very low metallicity 
similar to the extended  X-ray source in NGC 1068. The most likely
explanation is that another, pure continuum model is contributing at those 
energies. Ptak \etal\ suggested inverse Compton or accretion-driven point
sources, possibly X-ray binaries. 
They also suggest depletion by dust that will
help explain the low iron abundance. 
Given all  unknowns,
 we cannot be sure that most of the soft X-ray lines in NGC 1068 are
due to photoionized gas.

NGC 1068 differs significantly from starburst galaxies by showing large
EW low-ionization, H-like and He-like iron lines. 
The latter two are most probably due to high temperature
 photoionized gas and one must consider the consequences for 
production of other lines. 
Given solar metallicity, such gas must also 
produce strong argon, sulphur and silicon lines 
as well as some magnesium, neon and Fe-L emission (see Fig.2).
Most of these are in the 0.6--3 keV range where contribution from the
extended component is suspected. This makes
the simple hot plasma explanation even more questionable. Another important
difference is the intensity of the Fe-L lines. The Ptak \etal\ (1997)
observations of M82 and NGC~253, clearly show these lines to be weaker than computed
in solar metallicity hot plasma models. We find the lines, 
that are produced in our case 
 by recombination, not collisions, consistent with higher-than-solar
metallicity. 

\subsection{Reflection by the central torus}

An alternative explanation to the warm photoionized component was
recently proposed by 
  Iwasawa \etal\ (1997). These authors suggested that the 6.4 keV iron
line  
 originates in  the Compton thick ``walls'' of the central obscuring torus.
The  model  is different  from ours in two major ways. First,
because of the Compton thick  gas, there is more absorption, and
hence almost no reflection 
 below about 3 keV (see their Fig. 3). All
 the X-ray flux  below this energy is either due to 
the hot photoionized gas or the extended  emission.
 Additional manifestations
of the Compton thick medium are an extended low energy wing
 on the 6.4 keV line and  very strong 7.1 keV absorption. 

While we clearly identify
some flux excess at energies below 6.4 keV, with EW of about 100 eV, our fit
does not require any 7.1 keV absorption. However, the poor signal-to-noise
prevents us from reaching a firm conclusion on this point. As for the low-energy
wing, its EW 
  is of the same order as expected from 
 a relativisticly broadened nuclear line. 
If the continuum and line are both scattered, the
relativistic disk component would have the same equivalent width with
respect to that continuum, with $\sim 100$~eV EW 
in the line wing itself (cf Nandra \etal\ 1997),  as it does when it is directly
observed. Thus an
alternative explanation is that the X-ray mirror reflects both central
line and continuum. 

Finally, this explanation would require an even larger iron
composition since much of the produced 6.4 keV photons are absorbed by the
Compton thick gas.

The second major difference between our model and Iwasawa \etal\ is the 
origin of the nuclear mirror. This mirror 
is known to be extended and covers a sizable fraction of
the NLR. In our model, the  scattering of the optical
broad lines,  and optical-UV continuum, is due to both  warm and
hot components. The amount of reflection  is consistent with the required
column density and covering fraction (see Marshall \etal, 1993, for 
details). In the 
Iwasawa \etal\ (1997) model, only the hot component is extended, 
and contributes to the reflection, while the warm
X-ray gas occupies  a very small ($\sim 1$ pc) region.
Since very hot gas would 
broaden the reflected BLR Balmer lines beyond recognition, the model requires
another, yet unknown medium to explain the extended mirror. 

Iwasawa \etal\ (1997) also suggested a small redshift  of the
 He-like and H-like iron \Ka\ lines.
 We find a satisfactory fit for those lines with narrow
Gaussian at the systemic velocity.

\subsection{Low-Z fluorescence lines}

A comment on the intensity of some  soft X-ray metal lines is in order.
The fluorescence yield of low ionization species is a strongly increasing
function of the  atomic number. Thus, a fluorescence excitation 
of low-Z metals, such
as oxygen, neon and magnesium, is  thought to be negligible.
Our calculations clearly show that this is not the case. The reason is the steep
high energy continuum of AGN in general, and NGC~1068 continuum 
in particular. While the
yield is indeed small, the K-shell excitation energy of low-Z metals is much 
lower than in high Z species, 
and the much larger photon flux more than compensates for the lower yield.
We  predict that future X-ray experiments of sufficiently high resolution,
such as the grating instruments on AXAF and XMM, will discover relatively strong
fluorescence lines of low  ionization (i.e. lower than  Li-like ions) 
oxygen, neon, magnesium, silicon and sulphur.
  
Acknowledgements: We are grateful to Ian George, Richard Mushotzky, Paul
Nandra, Amiel Sternberg and Andy Ptak for useful comments and discussion.
This research is supported by a grant of the Israel Science
Foundation. 
We acknowledge the financial support of the Universities Space Research
Association (TJT).
The analysis was performed using {\sc XSELECT} (version 1.3)
and {\sc XSPEC} (version 9). 
This research made use of data 
obtained through the HEASARC on-line service, provided by NASA/GSFC.

\end{document}